# One dimensional surface profilometry by analyzing the Fresnel diffraction pattern from a step

**Osanloo, Soghra[1]; Darudi, Ahmad[2]**

[1] Physics Department, Institute for Advanced Studies in Basic Sciences (IASBS), 45195-1159, Zanjan, Iran
[2] Physics Department, Zanjan University, 45371-38791, Zanjan, Iran
[1] s_osanlou@iasbs.ac.ir
[2] darudi@znu.ac.ir

**Abstract:** When a coherent quasi-monochromatic light is reflected from a step, a diffraction pattern is formed that can be described by Fresnel-Kirchhoff integral and visibility of the fringes depends on the height of the step. In this paper, it is shown that the Fresnel diffraction from a step can be described by an interference-like formula. A relationship is derived between the visibility of the diffraction pattern from 1D step and the step height. Finally, a novel method is presented for 1D surface testing. The theoretical and experimental results are presented.
**OCIS codes:** (260.0260); (260.1960).

**Theoretical approach:** In Fig. 1, a cylindrical wavefront originating from a linear source strikes a 1D step of height $h$ [1]. The linear source and the edge of the step are perpendicular to the page. Applying Fresnel-Kirchhoff integral the diffracted amplitude and intensity can be calculated at an arbitrary point $P$ on a screen perpendicular to the reflected ray passing through point $P$. The amplitude and intensity at point $P$ depends on the location of $P_0$, the origin of the coordinate system used for the intensity calculation at point $P$ [1-3]. When $\theta_i = 0$, $\theta_i$ is incident angle, the amplitude at point $P$ is given by

$$u(x, y, z) = K' \left( \int_{-\infty}^{v_0} \exp(-i\pi v^2/2) dv + \exp(i\phi) \int_{v_0}^{\infty} \exp(-i\pi v^2/2) dv \right) \quad (1)$$

where $K' = \sqrt{\lambda/2(R+R')} K$, $\phi = 2hk$, $oo' = h$, $v = \sqrt{\frac{2}{\lambda}\left(\frac{1}{R}+\frac{1}{R'}\right)} x$, $v_0 = \sqrt{\frac{2}{\lambda}\left(\frac{1}{R}+\frac{1}{R'}\right)} x_0$ and $K$ is complex constant. $k$, $\lambda$ are wave number and wavelength of the incident light, respectively, $x_0$ is the distance of the point $P$ from the step edge.

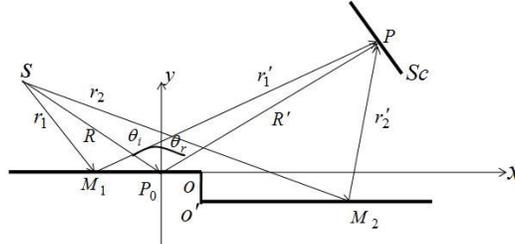

Fig 1. The geometry used to describe Fresnel diffraction from a step in reflection mode.

The integrals in Eq. (1) are Fresnel integrals that can be written

$$\int_{-\infty}^{v} \exp(-i\pi v^2/2) dv = a_1(x) \exp(i\phi_1(x)), \quad \int_{v}^{+\infty} \exp(-i\pi v^2/2) dv = a_2(x) \exp(i\phi_2(x)) \quad (2)$$

Substituting from Eq. (2) into Eq. (1) and considering that $I = u(x,y,z) u^*(x,y,z)$, I is Intensity and $u^*(x,y,z)$ is complex conjugate of $u(x,y,z)$, we obtain

$$I(x) = 2K'K'^*[1 + (2a_1(x)a_2(x)\sin(\phi/2))\cos(\pi/2 - ((\phi/2) - (\phi_1(x) - \phi_1(x))))] \quad (3)$$

The above equation is similar to the interference equation of two beams that their phase difference is $((\phi/2) - (\phi_1(x) - \phi_1(x)))$ and the fringes visibility in $x$ direction, $V(x)$, is $2a_1(x)a_2(x)\sin(\phi/2)$. So we can use the Fourier transform method [4], for analyzing the Fresnel diffraction fringes and calculate the visibility of diffraction pattern. Then, the visibility profile along the step is used to reconstruct the step height.

**Experimental works:** The scheme of the experimental setup is illustrated in Fig. 2. The step is constructed with putting the edge of a reference plate over a test plate. The expanded beam of a He-Ne laser perpendicularly strikes two plates. The diffracted light strikes a CCD after reflecting from beam splitter BS. The photograph in Fig. 3 is the recorded diffraction pattern.

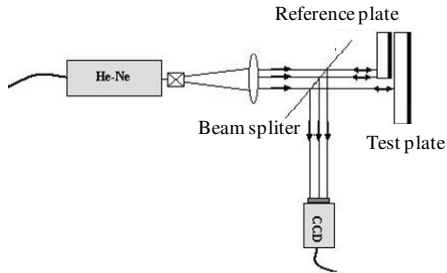
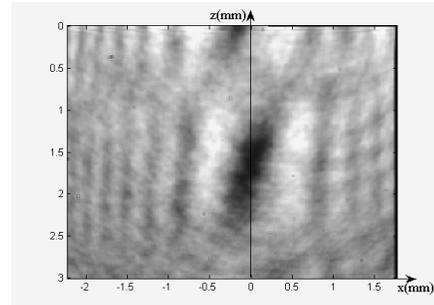

Fig.. 2. Sketch of the experimental setup.  Fig. 3. The Fresnel diffraction pattern of a step.

Using Fourier transform method, the visibility of the diffraction pattern is calculated for each horizontal line in Fig.3 and therefore visibility is known along solid line in Fig. 3. Then, with assumption of same reflectivity of upper and lower surfaces of the step and at x=0, from Eq.3 we have following relation between visibility and step height:

$$V(x=0, z) = \sin(\phi(x=0, z)/2) \qquad (4)$$

Fig. 4. shows the phase difference along the edge of the step.

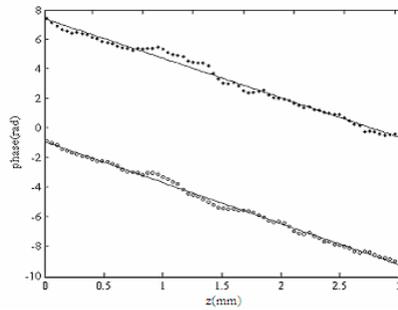

Fig. 4. The plots of the phase difference along the step edge derived by Fourier transform (solid circles) and phase shift (open circles).

In order to verify the reconstructed profile an alternative method should be used. The suitable method is phase shift fringe analysis. For this, the step has been attached to a PZT and few phase shifts have been introduced and phase shift algorithm is used for fringe analysis [5]. Fig. 4. shows the result of two methods, we expected nearly linear profile. The slopes of the two lines are nearly the same and their standard deviation is about .295 radian.

**Conclusion:** In this work we analytically derived a relationship between the visibility of 1D step diffraction fringes pattern and the step height. Therefore, a novel and simple method has been presented for 1D surface testing. In this method we used Fourier transform method. The height profile along the edge was reconstructed by analyzing the visibility of the Fresnel fringes pattern instead of the Intensity. In other words, current technique used fringe visibility rather than phase difference to calculate the step height since visibility is insensitive to source intensity fluctuation.